\documentstyle[11pt]{article}
\renewcommand{\thefootnote}{\fnsymbol{footnote}}
\newcommand{\newsection}{
\setcounter{equation}{0}
\section}

\def\appendix#1{
  \addtocounter{section}{1}
  \setcounter{equation}{0}
  \renewcommand{\thesection}{\Alph{section}}
 \section*{Appendix \thesection\protect\indent \parbox[t]{11.715cm} {#1}}
  \addcontentsline{toc}{section}{Appendix \thesection\ \ \ #1}
  }

\newcommand{\tr}[1]{\:{\rm Tr}\,#1}
\def\e{{\, e}\,}

\def\eop{\vspace*{\fill}\pagebreak}
\newcommand{\rf}[1]{(\ref{#1})}
\newcommand{\eq}[1]{eq.~(\ref{#1})}
\def\be{\begin{equation}}
\def\ee{\end{equation}}
\def\beq{\begin{equation}}
\def\eeq{\end{equation}}
\def\bea{\begin{eqnarray}}
\def\eea{\end{eqnarray}}
\def\LB{\left (}
\def\RB{\right )}

\def\cl{{\rm cl}}

\newcommand{\non}{\nonumber \\*}
\newcommand{\ie}{{\it i.e.}\ }

\newcommand{\ra}{\rightarrow}
\def\bl{\Bigl(}

\def\br{\Bigr)}

\begin{document}

\begin{titlepage}
\begin{flushright}
NBI--HE--97--09\\
ITEP--TH--09/97\\
hep-th/9703038\\
\end{flushright}
\vspace{.5cm}

\begin{center}
{\LARGE Towards a Non-perturbative Formulation of \\[.4cm]
IIB Superstrings by Matrix Models}\\
\vspace{1.2cm}
{\large A. Fayyazuddin${}^{{\rm a)}}$\footnote{E-mail: ansar@nbi.dk},
Y. Makeenko${}^{{\rm a) b)}}$%
\footnote{E-mail: makeenko@nbi.dk \ \ \ makeenko@vxitep.itep.ru},
P. Olesen${}^{{\rm a)}}$\footnote{E-mail: polesen@nbi.dk}, \\[.2cm]
D.J. Smith${}^{{\rm a)}}$\footnote{E-mail: D.Smith@nbi.dk},
and K. Zarembo${}^{{\rm b)}}$%
\footnote{E-mail: zarembo@vxitep.itep.ru} }\\
\vspace{24pt}
${}^{{\rm a)}}${\it The Niels Bohr Institute,}
\\ {\it  Blegdamsvej 17, DK 2100 Copenhagen \O, Denmark} \\[.2cm]
${}^{{\rm b)}}${\it Institute of Theoretical and Experimental Physics,}
\\ {\it B. Cheremushkinskaya 25, 117259 Moscow, Russia}
\end{center}
\vskip 0.9 cm
\begin{abstract}
We address the problem of a non-perturbative formulation of
superstring theory by means of the recently proposed matrix models.
For the model by Ishibashi, Kawai, Kitazawa and Tsuchiya (IKKT),
we perform one-loop calculation of the interaction between
operator-like solutions identified
with D-brane configurations of the type IIB superstring
(in particular, for parallel moving and rotated static p-branes).
Comparing to the superstring calculations,
we show that the matrix model reproduces the superstring
results only at large distances or small velocities,
corresponding to keeping
only the lowest mass closed string modes.
We propose a modification of the IKKT matrix model
introducing an integration over an additional
Hermitian matrix required to have positive definite eigenvalues, which is
similar to the square root of the metric in the
continuum Schild formulation of IIB superstrings.
We show that for this new matrix action the Nambu--Goto version of the
Green--Schwarz action is reproduced even at quantum level.
\end{abstract}

\end{titlepage}
\setcounter{page}{1}
\renewcommand{\thefootnote}{\arabic{footnote}}
\setcounter{footnote}{0}

\newsection{Introduction}

Recent interest in matrix-model formulation of superstrings
has been initiated by the proposal of Banks, Fischler, Shenker and
Susskind~\cite{BFSS96} that the non-perturbative dynamics of M~theory
is described by supersymmetric $n\times n$  matrix
quantum mechanics in the limit of large $n$.
This matrix model has been investigated in a number of subsequent
papers~\cite{grt,AB96,LM96,bss,Lif96}, and its operator-like classical
solutions were identified with D(irichlet) p-branes (for even $p$)
of the type IIA superstring theory.

Another matrix model has been proposed by Ishibashi, Kawai, Kitazawa and
Tsuchiya \cite{ikkt} (IKKT) for type IIB superstrings in ten dimensions.
The action of the model is defined by
\beq
S=\alpha\left( - \frac{1}{4}\tr[A_{\mu},A_{\nu}]^2
            -\frac{1}{2}\tr (\bar{\psi}
           \Gamma^{\mu}[A_{\mu},\psi])\right)+\beta n \,,
\label{action}
\eeq
where $A_{\mu}$ and $\psi_\alpha$ are $n \times n$ ($n\ra\infty$)
Hermitian bosonic and fermionic matrices, respectively.
The parameter $n$ is considered as a dynamical variable
which makes a crucial difference between the action~\rf{action}
and the one of (dimensionally reduced) ten-dimensional super Yang--Mills.
The action~\rf{action} is associated with the
IIB superstring in the Schild formalism (with fixed $\kappa$-symmetry):
\be
S_{\rm Schild}
=\int d^2\sigma \left(\alpha \Big(
\frac{1}{4\sqrt{g}}\{X^{\mu},X^{\nu}\}_{PB}^2
-\frac{i}{2}\bar{\psi}\Gamma^{\mu}\{X_{\mu},\psi\}_{PB}\Big)
+\beta \sqrt{g}\right),
\label{Saction}
\ee
where the commutators are substituted by the Poisson brackets.
Properties of the IKKT matrix model were further studied
in~\cite{Li96,cmz,fs}.%
\footnote{Another approach to the type IIB
superstring is discussed in~\cite{SS97,BS97}.}

The Dp-branes with odd $p$ of type IIB superstring theory
appear in the matrix model~\rf{action} as operator-like solutions of
the classical equations
\be
\left[ A^{\mu},\left[ A_\mu,A_\nu \right]\right]=0\,,~~~~~
\left[ A_\mu\,, (\Gamma^\mu\psi)_\alpha \right] =0\,.
\label{ce}
\ee
A general multi-brane solution has a block-diagonal form and is built
out of single p-branes. The solution associated with one p-brane
is given by
\be
A_\mu^\cl = \left(P_1,Q_1,\ldots,P_{\frac{p+1}{2}},
Q_{\frac{p+1}{2}},0,\ldots,0   \right),
~~~~~\psi_\alpha^\cl=0\,,
\label{DpPQ}
\ee
where $P$'s and $Q$'s form $(p+1)/2$ pairs of operators (infinite matrices)
obeying canonical commutation relation on a torus
associated with compactification (of large enough radii $L_a/2\pi$)
of the axes $0,\ldots,p$ so that
$L_aL_{a+1}/n^{2/(p+1)}$ (even $a$) is kept fixed as $n\ra\infty$.
This solution for D-string ($p=1$) was
constructed in~\cite{ikkt} in analogy with~\cite{BFSS96} and was extended
for $p\geq 3$ in~\cite{cmz,fs} in analogy with~\cite{bss}.
One of the arguments in favor of this construction is based
on the correct large-distance behavior of the one-loop
matrix model calculation of the interaction
between anti-parallel D-strings~\cite{ikkt} and higher
p-branes~\cite{cmz}.

In the present paper we continue the investigation of the
IKKT matrix model. We perform comprehensive one-loop calculation
of the interaction between two Dp-branes and compare the matrix-model
and superstring results.
For parallel moving p-branes
we demonstrate that the matrix model reproduces Bachas' superstring
result~\cite{Bachas} only at large distances or small velocities
keeping only the lowest mass closed string modes.
We propose a modification of the IKKT matrix model
introducing an additional (positive definite)
Hermitian matrix $Y^{ij}$, which is a dynamical variable to be
identified with $\sqrt{g}$ in \eq{Saction}.
The classical action has the form
\beq
S^{\rm cl}=\alpha\left( -\frac{1}{4}\tr Y^{-1}[A_{\mu},A_{\nu}]^2
            -\frac{1}{2}\tr (\bar{\psi}
           \Gamma^{\mu}[A_{\mu},\psi])\right)+\beta \tr Y
\label{Yaction}
\ee
and reduces
to the classical action of the non-abelian Born--Infeld (NBI) type:
\beq
S_{\rm nbi}^{\rm cl}=\sqrt{\alpha\beta} \tr \sqrt{-[A_{\mu},A_{\nu}]^2}
            -\frac{\alpha}{2}\tr (\bar{\psi}
           \Gamma^{\mu}[A_{\mu},\psi])\,,
\label{NBIaction}
\ee
using classical equation of motion for $Y$.
Moreover, we show that {\sl the Nambu--Goto version of the
Green--Schwarz action is reproduced even at the quantum level,
if one chooses appropriately the measure of integration over $Y$}. Our
results are therefore much more general than displayed in
eqs.~(\ref{Yaction}) and (\ref{NBIaction}).

In Sect.~2 we perform one-loop calculation of scattering of
parallel p-branes in the IKKT matrix model (as well as the interaction of
rotated static p-branes)
and compare to the superstring calculations. We also reproduce
previously known results for the anti-parallel p-branes using
the new technique.
In Sect.~3 we propose a modification of the IKKT matrix model
introducing integration over an additional Hermitian matrix $Y$, required to
have positive definite eigenvalues.
We show how the Nambu--Goto version of the
Green--Schwarz action is reproduced for the proposed model
at the quantum level. The results are discussed in Sect.~4.
Appendix~A contains the proof of the ${\cal N}=2$ supersymmetry of the
proposed matrix model for $n\rightarrow\infty$.

\newsection{Interaction of branes in the IKKT matrix model}

In the large-$n$ limit the matrices $A_\mu$ and $\psi_\alpha$ become
operators in a Hilbert space and the classical equations of
motion~\rf{ce} possess nontrivial solutions which possibly correspond
to solitonic states in type IIB superstring theory.
Among the solutions to \eq{ce}, the distinguished role is played by
the ones for which the field strength
 \begin{equation}\label{fmn}
 f_{\mu \nu }=i[A_{\mu },A_{\nu }]
 \end{equation}
 is proportional to the unit matrix.  Only
 these classical configurations can preserve half of the supersymmetries
 and thus can be interpreted as BPS states \cite{ikkt,bss}. This is why
 they are associated with D-branes of various dimensions.

 The solution which can be interpreted as D-brane of dimension $p$ has the
form
\be
A_\mu^\cl = \left(B_0,B_1,B_2,\ldots,B_p,0,\ldots,0   \right),
~~~~~\psi_\alpha^\cl=0\,,
\label{Dp}
\ee
where $B_0,\ldots,B_p$ are operators (infinite $n\times n$ matrices)
with the commutator
\be                        \label{bb}
[B_a,B_b]=-ig_{ab}{\bf 1}\,,
\ee
 and $a,b=0,\ldots,p$. Such solutions exist only for odd $p$,
otherwise one can find a linear combination of $B_{a}$'s, which commutes
with all other operators.
The solution corresponding to D-strings
($p=1$) was studied in~\cite{ikkt} and was generalized to $p=3$ and
$5$ in~\cite{cmz,fs}.

By a Lorentz transformation the skew--symmetric matrix
$g_{ab}$ can be brought to the canonical form
 \begin{equation}\label{stc}
 g_{ab}=\left(
 \begin{array}{ccccc}
 0 & -\omega_1 & & & \\
 \omega_1 &0& & & \\
  & &\ddots & & \\
 &&&0 & -\omega_{\frac{p+1}{2}}  \\
 &&&\omega_{\frac{p+1}{2}} &0 \\
 \end{array}
 \right).
 \end{equation}
 For such $g_{ab}$ the operators $B_a$ form a set of $l=(p+1)/2$ pairs of
 canonical variables and~\rf{Dp} coincides with~\rf{DpPQ}.
In the coordinate representation they can be represented by
 \begin{equation}\label{coordB}
 B_0=i\omega _1\partial _1,~~~~
 B_1=q_1,~~~~\ldots,~~~~
 B_{p-1}=i\omega _l\partial _l,~~~~
 B_p=q_l.
 \end{equation}
 The eigenvalues of the operators $B_a$ are uniformly distributed along
 the interval $[-L_a/2,L_a/2]$ of the real axis, where $L_a/2\pi$ are
 compactification radii. In fact, we can assume that the
 support of eigenvalues
 covers the whole real axis, because
 $L_a$ should scale in the large $n$ limit as $n^{\frac{1}{2l}}$
 \cite{BFSS96,ikkt,bss}. Each operator $B_a$ has $n^{\frac{1}{l}}$
 different eigenvalues, so the spacing between them,
 $L_an^{-\frac{1}{l}}$, scales as $n^{-\frac{1}{2l}}$.
 The product of the eigenvalue densities of
 the canonical conjugate variables $B_{2i-2}$ and $B_{2i-1}$
 is fixed by the Fourier transformation, so we get
 \begin{equation}\label{pdens}
\frac{n^{\frac{2}{l}}}{L_{2i-2}L_{2i-1}} =
                \frac{n^{\frac{1}{l}}}{2\pi\omega_i} \,.
\label{fuzz}
 \end{equation}

In this section we consider processes with interaction
between p-branes of this type.
We calculate first the interaction between two parallel
$p$-branes moving with constant velocity, then
the interaction between two p-branes rotated through some angle
and finally
the interaction between two anti-parallel p-branes (or brane-antibrane).

\subsection{Scattering of parallel p-branes}  \label{movingbranes}

The multi-brane configurations correspond to the solutions of the equations
of motion which have the block-diagonal form. Obviously, the matrices with
two identical blocks describe a pair of superimposed p-branes.
Let us first shift one of them by the distance $b/2$ and
the other by $-b/2$ along the $(p+2)$-th axis. This results
in the configuration of two parallel $p$-branes separated
along the $(p+2)$-th axis by the distance $b$ from each other.
Let us finally boost these stationary
$p$-branes along $(p+1)$-th axis
in opposite directions.
Using the block-diagonal construction,
the configuration of two parallel
$p$-branes moving with constant velocity, with impact parameter $b$,
is thus described by the following classical solution to \eq{ce}:
\begin{eqnarray}
A^{\rm cl}_0 & = & \left ( \begin{array}{cc} B_0 \cosh \epsilon & 0 \\
                0 & B_0 \cosh \epsilon \end{array} \right ), \non
A^{\rm cl}_a & = & \left ( \begin{array}{cc} B_a & 0 \\
                0 & B_a \end{array} \right ), ~~~ a=1,\ldots p, \non
A^{\rm cl}_{p+1} & = & \left ( \begin{array}{cc} B_0 \sinh \epsilon & 0 \\
                0 & -B_0 \sinh \epsilon \end{array} \right ), \non
A^{\rm cl}_{p+2} & = & \left ( \begin{array}{cc} \frac{b}{2} & 0 \\
                0 & -\frac{b}{2} \end{array} \right ), \non
A^{\rm cl}_i & = & 0, ~~~ i=p+3, \ldots 9\,.
\label{backgrv}
\end{eqnarray}
To simplify the calculation,
we have chosen the frame where the two $p$-branes have opposite velocity
$v$ and $-v$ along the $(p+1)$-th axis and
\begin{equation}
v = \tanh \epsilon\,.
\end{equation}

The interaction between the two $p$-branes to zeroth order in
string coupling constant is determined by the one-loop effective action
of the matrix model in the background~\rf{backgrv}, which
have the form~\cite{ikkt}
\be\label{seff}
W=\frac{1}{2}{\rm Tr}\ln(P^2\delta_{\mu
\nu}-2iF_{\mu\nu})-\frac{1}{4}{\rm Tr}\ln\LB (
P^2+\frac{i}{2}F_{\mu\nu}\Gamma^{\mu\nu})\LB\frac{1+\Gamma_{11}}{2}
\RB\RB
-{\rm Tr}\ln(P^2)\,,
\ee
after the Wick rotation to the Euclidean space.
The adjoint operators $P_{\mu}$ and $F_{\mu\nu}$ act on the space
of matrices and are defined by
\be
P_{\mu}=\left[A_{\mu }^\cl,\cdot\,\right],~~~~
 F_{\mu\nu}=i\left[\left[A_\mu^\cl,A_\nu^\cl\right],\cdot\,\right] .
\label{PF}
\ee

The block--diagonal form of the classical solution~\rf{backgrv} shows
that it is convenient to represent the matrices in adjoint
representation as $2\times 2$ matrices composed of $n\times n$
blocks.  At infinite $n$ these blocks become the operators acting in the
same Hilbert space as $B_a$.  In the coordinate representation~\rf{coordB}
they have the form of the functions of two sets of $l$ variables --- $q_i^1$
and $q_i^2$ --- and $B_a$ act on them as derivative (left and right) and
multiplication operators. From the definition of the adjoint operators,
we find that $P_{\mu}$'s act on Hermitian matrices as
\begin{eqnarray}\label{p'x}
P_0 \left ( \begin{array}{cc} X & Y \\ Y^{\dagger} & Z \end{array} \right )
   & = & i\omega_1(\partial_1^1+\partial_1^2)\cosh\epsilon
\left ( \begin{array}{cc} X & Y \\ Y^{\dagger} & Z \end{array} \right ), \non
P_1 \left ( \begin{array}{cc} X & Y \\ Y^{\dagger} & Z \end{array} \right )
   & = & (q_1^1-q_1^2)
\left ( \begin{array}{cc} X & Y \\ Y^{\dagger} & Z \end{array} \right ), \non
 & \ldots & \\
P_{p-1} \left ( \begin{array}{cc} X & Y \\ Y^{\dagger} & Z \end{array} \right )
   & = & i\omega_l(\partial_l^1+\partial_l^2)\cosh\epsilon
\left ( \begin{array}{cc} X & Y \\ Y^{\dagger} & Z \end{array} \right ), \non
P_p \left ( \begin{array}{cc} X & Y \\ Y^{\dagger} & Z \end{array} \right )
   & = & (q_l^1-q_l^2)
\left ( \begin{array}{cc} X & Y \\ Y^{\dagger} & Z \end{array} \right ), \non
P_{p+1} \left ( \begin{array}{cc} X & Y \\ Y^{\dagger} & Z \end{array} \right )
   & = & i\omega_1\sinh\epsilon
   \left ( \begin{array}{cc} (\partial_1^1+\partial_1^2)X &
   (\partial_1^1-\partial_1^2)Y \\ -(\partial_1^1-\partial_1^2)Y^{\dagger} &
   -(\partial_1^1+\partial_1^2)Z \end{array} \right ), \non
P_{p+2} \left ( \begin{array}{cc} X & Y \\ Y^{\dagger} & Z \end{array} \right )
   & = & b
   \left ( \begin{array}{cc} 0 & Y \\ -Y^{\dagger} & 0 \end{array} \right ).
\end{eqnarray}

The only non-zero component of the field strength in the adjoint
representation is $F_{1\,p+1}$ which acts as
\begin{eqnarray}\label{fp-1p+1}
F_{1\,p+1}\left ( \begin{array}{cc} X &
      Y\\ Y^{\dagger} & Z \end{array} \right )
   & = & 2\omega_1\sinh\epsilon
   \left ( \begin{array}{cc} 0 & Y \\ -Y^{\dagger} & 0 \end{array} \right ).
\end{eqnarray}
This operator acts non-trivially only on $Y$ and $Y^{\dagger}$.
The effective action~\rf{seff} vanishes for $F_{\mu \nu }=0$~\cite{ikkt},
consequently we can trace only the action of $P^2$ on $Y$. The
contributions of the eigenvalues, corresponding to the eigenfunctions
with nonzero $X$ and $Z$, are mutually cancelled.

Using the notation
\begin{eqnarray}
\label{q12q+-}
q_i^{\pm} & = & \frac{1}{\sqrt{2}}(q_i^1 \pm q_i^2) \non
\partial_i^{\pm} & = & \frac{1}{\sqrt{2}}(\partial_i^1 \pm \partial_i^2)\,,
\end{eqnarray}
we write the action of the operator $P^2$ as
\begin{equation}
P^2Y = \left\{
b^2 + 2\sum_{i=2}^{l}\left [ (q_i^-)^2-(\omega_i\partial_i^+)^2 \right ]
   - 2\omega_1^2\cosh^2\epsilon~(\partial_1^+)^2 + 2(q_1^-)^2 -
   2\omega_1^2\sinh^2\epsilon~(\partial_1^-)^2\right\} Y\,.
\end{equation}
This expression is calculated in Euclidean space after a Wick rotation. We
can clearly see that the last two terms that
depend on $q_1^-$ and $\partial _1^-$
give a Hamiltonian of harmonic oscillator, while all
$\partial_i^+$, and $q_i^-$ for $i=2,\ldots,l$ enter without their
conjugate variables and can be simultaneously diagonalized with
$P^2$.
Thus the eigenvalues of $P^2$ are given by
\begin{equation}
E_{{\bf q},{\bf p},k} = b^2 + 2\sum_{i=2}^{l}(q_i^2+p_i^2) +
 2\cosh^2\epsilon p_1^2 + 4\omega_1\sinh\epsilon~(k+\frac{1}{2}) \,,
\label{spec}
\end{equation}
where $q_i$ and $p_i$ are the eigenvalues of $q_i^-$ and
$i\omega_i\partial_i^+$, respectively.

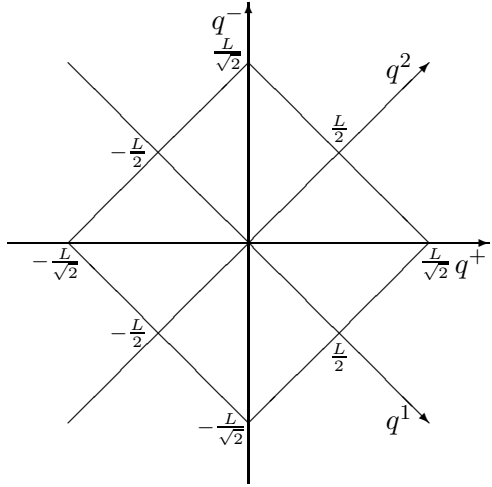
\begin{figure}[tbp]
\unitlength .8mm
\linethickness{0.6pt}
\begin{picture}(80.00,100.00)(-65,-10)
\put(0.00,40.00){\vector(1,0){80.00}}
\put(40.00,0.00){\vector(0,1){80.00}}
\put(10.00,40.00){\line(1,1){30.00}}
\put(40.00,70.00){\line(1,-1){30.00}}
\put(70.00,40.00){\line(-1,-1){30.00}}
\put(40.00,10.00){\line(-1,1){30.00}}
\put(10.00,70.00){\vector(1,-1){60.00}}
\put(10.00,10.00){\vector(1,1){60.00}}
\put(36.50,77.00){\makebox(0,0)[cc]{$q^-$}}
\put(77.00,36.50){\makebox(0,0)[cc]{$q^+$}}
\put(65.00,10.50){\makebox(0,0)[cc]{$q^1$}}
\put(65.00,69.00){\makebox(0,0)[cc]{$q^2$}}
\put(55.00,59.00){\makebox(0,0)[cc]{${\scriptstyle \frac{L}{2}}$}}
\put(20.00,25.00){\makebox(0,0)[cc]{${\scriptstyle -\frac{L}{2}}$}}
\put(20.00,55.00){\makebox(0,0)[cc]{${\scriptstyle -\frac{L}{2}}$}}
\put(55.00,21.00){\makebox(0,0)[cc]{${\scriptstyle \frac{L}{2}}$}}
\put(8.00,36.50){\makebox(0,0)[cc]{${\scriptstyle -\frac{L}{\sqrt{2}}}$}}
\put(71.00,36.50){\makebox(0,0)[cc]{${\scriptstyle \frac{L}{\sqrt{2}}}$}}
\put(35.50,9.00){\makebox(0,0)[cc]{${\scriptstyle -\frac{L}{\sqrt{2}}}$}}
\put(36.50,71.30){\makebox(0,0)[cc]{${\scriptstyle \frac{L}{\sqrt{2}}}$}}
\end{picture}
\caption {The region of integration in the $q^+,q^-$-plane.}
\label{square}
\end{figure}

 According to the discussion at the beginning of this section, the
 eigenvalues of the operators $q_i^1$, $q_i^2$ are distributed
 from $-L_{2i-1}/2$ to $L_{2i-1}/2$ with the constant density
 $n^{1/l}/L_{2i-1}$.  This is not true for the eigenvalues of
 $q_i^-$, $q_i^+$ because the integration region changes
 under the change of variables~\rf{q12q+-},
 as shown in fig.~\ref{square}.
 As a result, the density of the eigenvalues $q_i$ in \rf{spec} decreases
 from $\sqrt{2}n^{1/l}/L_{2i-1}$ at the origin to zero at $\pm
 L_{2i-1}/\sqrt{2}$. The scale of the variation is however of
 order $L$, which is negligible in the large--$n$ limit.
 For convergent integrals the distribution of $q_i$ and $p_i$
 can be taken to be uniform and equal to its value at the origin,
 $\sqrt{2}n^{1/l}/L_{2i-1}$ for $q_i$ and $\sqrt{2}n^{1/l}/L_{2i-2}$ for
 $p_i$.

Now we can utilize the results of Ref.~\cite{ikkt} to bring
the one-loop effective action for the given background to the form
 \begin{eqnarray}\label{seff1}
 W&=&\prod_{a\ne 1}
 \left(
 \frac{\sqrt{2}n^{\frac{1}{l}}}{L_a}
 \right)
 \int d^{l-1}q\,d^{l}p\,
 \sum_{k=0}^{\infty }\left[
 \ln\left(1-\frac{16\omega^2 _1\sinh^2 \epsilon}
 {E_{{\bf q},{\bf p},k}^2}\right)
 \vphantom{-\frac{1}{2}\sum_{
 \begin{array}{c}
 \scriptstyle
 s_1,\ldots,s_5=\pm 1\\[-2.5mm]
 \scriptstyle
 s_1\ldots s_5=1
 \end{array}
 }\ln\left(1-\frac{2\omega _1s_1}{E_{{\bf q},{\bf p},k}}
 \right)}
 \right.\non&&\left.
 -\,\frac{1}{2}\sum_{
 \begin{array}{c}
 \scriptstyle
 s_1,\ldots,s_5=\pm 1\\[-2.5mm]
 \scriptstyle
 s_1\ldots s_5=1
 \end{array}
 }\ln\left(1-\frac{2s_1\omega _1\sinh\epsilon}
 {E_{{\bf q},{\bf p},k}}
 \right)
 \right].
 \end{eqnarray}
 The last term which originates from the integration over fermions
 can be rewritten as
 \be
 \vphantom{\frac{1}{2}\sum_{
 \begin{array}{c}
 \scriptstyle
 s_1,\ldots,s_5=\pm 1\\[-2.5mm]
 \scriptstyle
 s_1\ldots s_5=1
 \end{array}
 }\ln\left(1-\frac{2\omega _1s_1}{E_{{\bf q},{\bf p},k}}
 \right)}
 \,\frac{1}{2}\sum_{
 \begin{array}{c}
 \scriptstyle
 s_1,\ldots,s_5=\pm 1\\[-2.5mm]
 \scriptstyle
 s_1\ldots s_5=1
 \end{array}
 }\ln\left(1-\frac{2s_1\omega _1\sinh\epsilon}
 {E_{{\bf q},{\bf p},k}}
 \right) = 4
 \ln\left(1-\frac{4\omega _1^2\sinh^2\epsilon}
 {E^2_{{\bf q},{\bf p},k}}
 \right).
 \ee

 It is convenient to represent the logarithms in \rf{seff1} in the form
 of the integrals over a ``proper time'' $s$:
   \begin{equation}\label{log1}
 \ln \frac uv =\int\limits_0^{\infty}\frac{ds}{s}\left(\e^{-vs}
- \e^{-us} \right).
 \end{equation}
 The sum over $k$ and the integrals over $q$ and $p$ then can be evaluated
 using the formula
 \begin{equation}\label{log2}
 \int d^{l-1}q\,d^{l}p\, \sum_{k=0}^{\infty }\e^{-sE_{{\bf q},{\bf p},k}}
 =\left(\frac{\pi
 }{2s}\right)^{\frac{p}{2}}\frac{\e^{-b^2s}}{2\cosh\epsilon\,
 \sinh\left(2\omega _1s\sinh\epsilon\right)}.
 \end{equation}
 We finally obtain the following form
\begin{equation}
 W=-n^{\frac{2p}{p+1}}
 \prod_{a\ne 1}L_a^{-1} \int_0^{\infty}
        \frac{ds}{s}\left ( \frac{\pi}{s} \right )^{\frac{p}{2}}
       e^{-b^2s} \frac{(\cosh(4\omega_1s\sinh\epsilon) -
        4\cosh(2\omega_1s\sinh\epsilon) + 3)}
        {\cosh\epsilon~\sinh(2\omega_1s\sinh\epsilon)}\,.
\label{otvetv}
\end{equation}

Defining
\begin{equation}
V_{p+1} = \prod_{a=0}^{p} L_a
\end{equation}
and substituting (cf. \eq{fuzz})
\begin{eqnarray}
n^{\frac{1}{l}} & = &
\frac{L_{2i-2}L_{2i-1}}{2\pi\omega_i}\,, \non n & = &
V_{p+1}\prod_{i=1}^{l}\frac{1}{2\pi \omega_i} \,,
\end{eqnarray}
we now Wick rotate back to Minkowski space-time to obtain
\begin{equation}
W=-i\frac{V_p}{(2\pi)^p}\omega_1\prod_{i=1}^{l}\frac{1}{\omega_i^2}
        \int_0^{\infty}
        \frac{ds}{s}\left ( \frac{\pi}{s} \right )^{\frac{p}{2}}
        e^{-b^2s}\frac{(\cos(4\omega_1s\sinh\epsilon) -
        4\cos(2\omega_1s\sinh\epsilon) + 3)}
        {\cosh\epsilon~\sin(2\omega_1s\sinh\epsilon)}
\label{wicky}
\end{equation}
where
\begin{equation}
V_p=\prod_{a=1}^{p}L_a\,.
\end{equation}
It should also be noted that $\omega_i \sim \alpha'$ from dimensional
analysis. We take
\begin{equation}
\omega_i = 2\pi\alpha'\,,~~~~~
\label{fuzz2}
\end{equation}
for {\sl all} $i$, which, as we shall now show, is the correct normalization
to get agreement with supergravity.

We can compare the result we get from the IKKT model to that
of Bachas \cite{Bachas} which is exact in $b,v,\alpha '$.
It is clear that the
IKKT result does not agree with Bachas' calculation, for instance
a comparison of the absorptive parts shows that they do not have the
same poles (see below).  There is, however, a regime in which the two results
are identical.  This is the regime in which supergravity, or alternatively
the lightest closed string modes dominate the interactions.  Thus
this is a low-energy long distance approximation.  This regime is
characterized by $b^{2} \gg\alpha '$.  Bachas' expression for the
one-loop effective action can be written in the more concise form
\cite{grgu}:
\begin{equation}
F = V_{p}(8\pi^{2}\alpha ')^{-p/2}\int_{0}^{\infty}
d\tau~\tau^{-4 +p/2}e^{-b^{2}/2\pi\alpha '{\tau}}
\left[ e^{-\pi\tau /12}\prod_{n=1}^{\infty}
(1-e^{-2\pi\tau n})\right]^{-9}\frac{(\theta_{1}(-i\epsilon /\pi\mid i\tau))^4}
{\theta_{1}(-i2\epsilon/\pi \mid i\tau)}.
\label{thetat}
\end{equation}
In the $b^{2} \gg\alpha '$ limit the integral is dominated by large
values of $\tau$ thus we can expand the theta functions in $q =e^{-\pi\tau}$
to get:
\begin{equation}
\theta_{1}( z\mid i\tau) = -2q^{1/4}\sin(\pi z) + \mbox{higher orders in $q$}.
\end{equation}
Keeping the lowest order in $q$ yields
\begin{equation}
F = -4iV_{p}(8\pi^{2}\alpha ')^{-p/2}\;
\frac{\sinh^{3}\epsilon}{\cosh \epsilon}
\int_{0}^{\infty}
d\tau~\tau^{-4 +p/2}e^{-b^{2}/2\pi\alpha '{\tau}}\,.
\end{equation}
By changing the integration variable to $t= 1/\tau$ one can re-write the
expression as
\bea
F& =& -4iV_{p}(8\pi^{2}\alpha ')^{-p/2} \;
\frac{\sinh^{3}\epsilon}{\cosh \epsilon}
\int_{0}^{\infty}
dt~t^{2-p/2}e^{-b^{2}t/2\pi\alpha '} \non
&=& -4i V_p (4\pi)^{-p/2}\:
\frac{\sinh^{3}\epsilon}{\cosh \epsilon}\:
\Gamma\left( \frac{6-p}{2}\right)\:
\frac{(2\pi\alpha')^{3-p}}{b^{6-p}}\,.
\eea
One can compare this to the expression we have obtained above using the
matrix model.  If we perform a similar approximation in which $b^2\gg\alpha '$,
we find that our expression is dominated by small $s$.
We obtain an identical expression to
Bachas in the above approximation scheme. Also, we mention that we get
complete agreement between the matrix model and superstrings for small velocity
and any $b\neq 0$. This is due to the complete cancellation of all the factors
containing exp$(-\pi\tau)$ in eq. (\ref{thetat}). However, the next order
term in the expansion of the small velocities does not agree.

One might still ask how much the results from the matrix model deviates from
the results obtained by Bachas. Due to the complicated integrals over theta
functions, it is not so easy in general to give a quantitative estimate
of these deviations. However, if we compare the imaginary part of the
phase shift\footnote{The real and imaginary parts are defined by
$W=i($Re $\delta+i~$Im$~\delta)$.},
computed for D-branes by Bachas~\cite{Bachas}, with the same
quantity computed from the matrix model, it is quite
easy to see that in the limit where the velocity approaches light velocity,
there is a very large physical difference between the results.
The matrix result (\ref{wicky}) has an imaginary part corresponding to
poles coming from the trigonometric functions at the values
\begin{equation}
s_k=\frac{\pi k}{2\omega_1 \sinh\epsilon},~{\rm for}~k=1,3,5,...~.
\label{im1}
\end{equation}
Computing the residue, we see (normalizing the parameters $\omega_i$
as in eq.~(\ref{fuzz2})) that for small velocities we get a
result which is the same as the one obtained by
Bachas\footnote{The units are $\alpha'=1/2$. Also, the velocity used here is
one half of the velocity used by Bachas.},
\begin{equation}
{\rm Im}~\delta\approx \frac{8~V_p}{(2\pi^3)^{p/2}}~v^{p/2}e^{-b^2/2v}
\label{immy}
\end{equation}
to leading order. It should be emphasized that this result is an independent
check of the matrix model versus superstring calculations. This is because
when we compute the large distance Re $\delta$, the relevant region of
integration is $s\rightarrow 0$, whereas for Im $\delta$ the relevant $s$
is given by the position of the lowest pole. The result (\ref{immy})
also implies that the normalization (\ref{fuzz}) is correct for the
imaginary part of the phase shift.

However, for a large velocity the results differ. Introducing \cite{Bachas}
\begin{equation}
\epsilon\approx~\ln(1-v)/2\approx~\ln(\bar{s}/{\cal M}_p^2)\gg 1,
\label{im2}
\end{equation}
where $\bar{s}$ stands for usual Mandelstam's variable and
${\cal M}_p$ is the mass of the $p$-brane, we get
\begin{equation}
{\rm Im}~\delta\approx {\rm const}\left(\frac{\bar{s}}{{\cal M}_p^2}\right)
^{p/2-1}\exp(-b^2{\cal M}_p^2/\bar{s}).
\label{im3}
\end{equation}
The main difference with the D-brane case is that these behave like black
absorptive disks of logarithmically growing area, $b_{cr}^2\sim
\ln(\bar{s}/{\cal M}_p^2)$, whereas from the matrix model these black disks
are much larger, corresponding to $b_{cr}^2\sim \bar{s}/{\cal M}_p^2$.

 \subsection{Rotated branes} \label{crossedbranes}

 The configuration with two rotated p-branes can be obtained
 from the block-diagonal matrix with two identical blocks describing
 a pair of superimposed p-branes quite similarly to
 Subsect.~\rf{movingbranes}.
 Shifting along the $(p+2)$-th axis by the distance $b$
 from each other and rotating in opposite directions in $(p,p+1)$ plane
through the angle $\theta /2$, we obtain the
configuration of two rotated branes
 \begin{eqnarray}\label{backgr}
 A_{a}^\cl&=&\left(
 \begin{array}{cc}
 B_a & 0 \\
 0 & B_a \\
 \end{array}
 \right),~~~~a=0,\ldots,p-1,\non
 A_{p}^\cl&=&\left(
 \begin{array}{cc}
 B_p\cos\frac{\theta }{2} & 0 \\
 0 & B_p\cos\frac{\theta }{2} \\
 \end{array}
 \right),      \non
 A_{p+1}^\cl&=&\left(
 \begin{array}{cc}
 B_p\sin\frac{\theta }{2} & 0 \\
 0 & -B_p\sin\frac{\theta }{2} \\
 \end{array}
 \right),          \non
 A_{p+2}^\cl&=&\left(
 \begin{array}{cc}
 \frac{b}{2} & 0 \\
 0 & -\frac{b}{2} \\
 \end{array}
 \right),\non
 A_{i }^\cl&=&0,~~~~i =p+3,\ldots,9 \,.
 \end{eqnarray}
This looks just like an analytic continuation of \eq{backgrv}.
Therefore all the formulas of this subsection are quite
similar to those of the previous one.

 The interaction between these rotated p-branes to zeroth order in
 string coupling constant is determined by the one-loop effective
 action~\rf{seff} in the background~\rf{backgr}.
 Repeating the calculation, we arrive at the Hamiltonian
 which has the spectrum (cf.~\rf{spec})
 \begin{equation}\label{specc}
 E_{{\bf q},{\bf
 p},k}=b^2+2\sum_{i=1}^{l-1}(q_i^2+p_i^2)+2\cos^2\frac{\theta }{2}\,~q_l^2
 +4\omega _l\sin^2\frac{\theta }{2}\,~(k+\frac{1}{2})\,.
 \end{equation}

The final result for the interaction between two rotated p-branes,
which are separated by the distance $b$ we represent in the form
 \begin{eqnarray}\label{otvet}
 W&=&-4n^{\frac{2p}{p+1}}
 \frac{1}{\cos\frac{\theta}{2}}\prod_{a\ne p-1}L_a^{-1}
 \non&&\times
 \int\limits_{0}^{\infty }\frac{ds}{s}
 \left(\frac{\pi}{s}\right)^{\frac{p}{2}}\,
 \e^{-b^2s}\tanh\left(\omega _{\frac{p+1}{2}}s\sin\frac{\theta }{2}\right)
 \sinh^2\left(\omega _{\frac{p+1}{2}}s\sin\frac{\theta }{2}\right).
 \end{eqnarray}
It can be obtained from~\rf{otvetv} substituting $\epsilon=i\theta/2$.

 For large separation between branes we find, using the equality
 \rf{pdens}:
 \begin{equation}\label{answ}
 W=-\Gamma\left(\frac{6-p}{2}\right)\,
 \frac{4V_p}{(4\pi )^{\frac{p}{2}}}
 \,\omega^4_{\frac{p+1}{2}}\prod_{i}\frac{1}{\omega _i^2}
 \,\frac{\sin^3\frac{\theta }{2}}{\cos\frac{\theta
 }{2}}\,\frac{1}{b^{6-p}} +O\left(\frac{1}{b^{8-p}}\right).
 \end{equation}
 This expression correctly
 reproduces the supergravity result for the angular and distance
 dependence of the interaction energy between two rotated p-branes.
 An analogous formula for $p=1$ is first obtained in~\cite{ikkt}.

 \subsection{Anti-parallel branes}

 The interaction potential for two anti-parallel D-strings in
 the IKKT matrix model
 was calculated in~\cite{ikkt}. This calculation has been generalized
 to p-branes of arbitrary dimension~\cite{cmz}. For completeness, we
 reproduce these results here using the same techniques as
 in the two previous subsections.

 The classical solution describing
 anti-parallel p-branes at the distance $b$ from each other are
 represented by block--diagonal matrices
 \begin{eqnarray}\label{apll}
 A_{a}^\cl&=&\left(
 \begin{array}{cc}
 B_a & 0 \\
 0 & B'_a \\
 \end{array}
 \right),~~~~a=0,\ldots,p\non \label{backgr1}
 A_{p+1}^\cl&=&\left(
 \begin{array}{cc}
 b/2 & 0 \\
 0 & -b/2 \\
 \end{array}
 \right),\non    \label{backgr2}
 A_{i }^\cl&=&0,~~~~i =p+2,\ldots,9,
 \end{eqnarray}
 where $B_a$ and $B'_a$ obey the commutation relations
 \begin{equation}\label{qp}
 [B_a,B_b]=-ig_{ab}{\bf 1},~~~~
 [B'_a,B'_b]=ig_{ab}{\bf 1}.
 \end{equation}
 The matrix $g_{ab}$ can be taken in the same form as in eq.~\rf{stc}.
 Thus we put
 \begin{eqnarray}\label{coordBB'}
 &&B_0=i\omega _1\partial _1,~~~~
 B_1=q_1,~~~~\ldots,~~~~
 B_{p-1}=i\omega _l\partial _l,~~~~
 B_p=q_l,\non
 &&B'_0=i\omega _1\partial _1,~~~~
 B'_1=-q_1,~~~~\ldots,~~~~
 B'_{p-1}=i\omega _l\partial _l,~~~~
 B'_p=-q_l,
 \end{eqnarray}
 where $l=(p+1)/2$.

 To calculate the one-loop effective action in the background~\rf{apll},
 we perform the same steps as in Subsec.~\ref{movingbranes}.
 First, we find all nonzero components of the adjoint operators
 defined by eq.~\rf{PF}:
 \begin{eqnarray}\label{p'xap}
 P_0\left(
 \begin{array}{cc}
 X & Y \\
 Y^{\dagger} & Z \\
 \end{array}
 \right)
 &=&i\omega _1(\partial _1^1+\partial _1^2)
 \left(\begin{array}{cc}
 X & Y \\
 Y^{\dagger} & Z \\
 \end{array}
 \right),       \non
 P_1\left(
 \begin{array}{cc}
 X & Y \\
 Y^{\dagger} & Z \\
 \end{array}
 \right)
 &=&\left(\begin{array}{cc}
 (q_1^1-q_1^2)X & (q_1^1+q_1^2)Y \\
 -(q_1^1+q_1^2)Y^{\dagger} & -(q_1^1-q_1^2)Z \\
 \end{array}
 \right),      \non
 &\cdots&      \non
 P_{p-1}\left(
 \begin{array}{cc}
 X & Y \\
 Y^{\dagger} & Z \\
 \end{array}
 \right)
 &=&i\omega _l(\partial _l^1+\partial _l^2)
 \left(\begin{array}{cc}
 X & Y \\
 Y^{\dagger} & Z \\
 \end{array}
 \right),       \non
 P_p\left(
 \begin{array}{cc}
 X & Y \\
 Y^{\dagger} & Z \\
 \end{array}
 \right)
 &=&\left(\begin{array}{cc}
 (q_l^1-q_l^2)X & (q_l^1+q_l^2)Y \\
 -(q_l^1+q_l^2)Y^{\dagger} & -(q_l^1-q_l^2)Z \\
 \end{array}
 \right),      \non
 P_{p+1}\left(
 \begin{array}{cc}
 X & Y \\
 Y^{\dagger} & Z \\
 \end{array}
 \right)
 &=&b
 \left(\begin{array}{cc}
 0 & Y \\
 -Y^{\dagger} & 0 \\
 \end{array}
 \right)
 \end{eqnarray}
 and
 \begin{equation}\label{fabap}
 F_{2i-2\,2i-1}\left(
 \begin{array}{cc}
 X & Y \\
 Y^{\dagger} & Z \\
 \end{array}
 \right)
 =-2\omega _i\,
 \left(\begin{array}{cc}
 0 & Y \\
 -Y^{\dagger} & 0 \\
 \end{array}
 \right),~~~~i=1,\ldots,l.
 \end{equation}
 Again $F_{\mu \nu }$ acts only on $Y$; the operator $P^2$ in this
 subspace has the form
 \begin{equation}\label{p2yap}
 P^2Y=\left\{b^2+2\sum_{i=1}^{l}\left[(q_i^+)^2-\omega _i^2(\partial
 _i^+)^2\right]\right\} Y\,,
 \end{equation}
 which coincides with the Hamiltonian of the $l$-dimensional harmonic
 oscillator. Therefore the eigenvalues of $P^2$ are labelled by the set of
 $l$ positive integers and are given by
 \begin{equation}\label{eigenap}
 E_{\bf k}=4\sum_{i=1}^{l}\omega _i
 \left(k_i+\frac{1}{2}\right)+b^2.
 \end{equation}
 The form of the adjoint field strength \rf{fabap} allows to use the
 representation for the effective action found in \cite{ikkt}, the same
 as in eq.~\rf{seff1}:
 \begin{equation}\label{seff1ap}
 W=n\sum_{\bf k}\left[\sum_{i}
 \ln\left(1-\frac{16\omega^2 _i}{E_{\bf k}^2}\right) -\frac{1}{2}\sum_{
 \begin{array}{c}
 \scriptstyle
 s_1,\ldots,s_5=\pm 1\\[-2.5mm]
 \scriptstyle
 s_1\ldots s_5=1
 \end{array}
 }\ln\left(1-\frac{2\sum\limits_{i}\omega _is_i}{E_{\bf k}}\right)\right].
 \end{equation}
 Neither the quantities $q_i^-$, nor their conjugate variables enter
 \rf{p2yap}, so the trace over them gives an overall factor of $n$.

 Using the same proper time representation for logarithms as in
 eq.~\rf{log1} and the equality
 \begin{equation}\label{log2ap}
 \sum_{\bf k}\e^{-sE_{\bf k}}=\frac{\e^{-b^2s}}{\prod\limits_i
 2\sinh 2\omega_i s},
 \end{equation}
 we obtain for the effective action:
 \begin{equation}\label{otvetap}
 W=-2n\int\limits_{0}^{\infty }\frac{ds}{s}\,\e^{-b^2s}
 \left[\sum_{i}\bl\cosh 4\omega _is-1\br-4\bl\prod_{i}\cosh
 2\omega _is-1\br\right]\prod_{i}\frac{1}{2\sinh 2\omega _is}.
 \end{equation}
 For large separation between the branes, this result reduces to
 \begin{equation}\label{1/bap}
 W=-\,\frac{1}{16}\,n\:\Gamma\left(\frac{7-p}{2}\right)\,
 \left[2\sum_{i}\omega _i^4-\left(\sum_{i}\omega _i^2\right)^2\right]
 \prod_{i}\omega _i^{-1}\left(\frac{2}{b}\right)^{7-p}
 +O\left(\frac{1}{b^{9-p}}\right).
\end{equation}
Equations~\rf{otvetap} and \rf{1/bap} coincide with the previous
results~\cite{ikkt,cmz} obtained by a slightly different technique.

These results of the matrix models are to be compared to
the superstring calculations which are given in the open-string
language by the annulus diagram. The superstring result
for the interaction
between anti-parallel Dp-branes reads~\cite{Pol95,grgu}
\be
W=-V_{p+1}\int_0^\infty \frac{dt}{t}
\frac{1}{\left(8\pi^2 \alpha' t\right)^\frac{p+1}{2}}
\e^{-{b^2 t}/{2\pi\alpha'}}
q^{-1}\frac{\prod_{n=1}^\infty (1-q^{2n-1})^8}
{\prod_{n=1}^\infty (1-q^{2n})^8}
\label{superstring}
\ee
with $q=\e^{-\pi t}$.
We see that the superstrings and matrix-model answers
agree only at large distances%
\footnote{Tseytlin~\cite{Tse97} has conjectured
an alternative interpretation of the classical solutions
in the IKKT matrix model as D-branes with magnetic field,
in analogy with previous work~\cite{LM96}
on the matrix model~\cite{BFSS96}. We do not
discuss such a point of view in this paper.}
quite similarly to the cases of moving and rotated branes.
This suggests to modify the matrix model to better reproduce
the superstring calculation.

\newsection{The NBI-type matrix model of IIB superstring}

In the IKKT model the matrix size $n$ is considered as a
dynamical variable. The partition function includes for
this reason a summation over $n$:
\beq
Z= \sum_{n=1}^{\infty} \int {\cal D}A \,{\cal D}\psi \, \e^{-S}\,,
\label{main}
\eeq
where the action is given by \eq{action}.
This construction is proposed in~\cite{ikkt} as
the matrix-model analog of the Schild formulation
of type IIB superstring given by the path integral
\beq
{\cal Z} = \int\, {\cal D} \sqrt{g}\, {\cal D}  X\,
{\cal  D} \psi \,\e^{- S_{\rm Schild}}
\label{Smain}
\eeq
with the action~\rf{Saction}.

In this section we propose a modification of the IKKT matrix model
which appears to be a more analogous to \eq{Smain}, and which reproduces
the Nambu--Goto version of the Green--Schwarz superstring action
after the integration over the introduced additional
Hermitian matrix $Y^{ij}$ with positive definite eigenvalues, which is
analogous to $\sqrt{g}$ in \eq{Smain}.
The classical action has the form~\rf{Yaction}, which yields
the following classical equation of motion for the $Y$-field:
\be
\frac{\alpha}{4}
\left( Y^{-1}[A_{\mu},A_{\nu}]^2 Y^{-1}  \right)_{ij} +
\beta\delta_{ij}=0 \,,
\ee
whose solution reads
\be
Y= \frac{1}{2}\sqrt{\frac\alpha\beta}\sqrt{ -[A_{\mu},A_{\nu}]^2 } \,.
\label{Ycl}
\ee
Here $-[A_{\mu},A_{\nu}]^2$ is positive definite, since the commutator is
anti-hermitian. The square root in \rf{Ycl}
is unique, provided $Y$ is positive definite which is our case.
After the substitution~\rf{Ycl}, the classical action~\rf{Yaction}
reduces to the classical action of the NBI (non-abelian Born-Infeld)
type~\rf{NBIaction}.

In this section we show that even at the quantum level, it is possible to
modify the classical action (\ref{Yaction}) such that we obtain the
Nambu-Goto version of the Green-Schwarz type IIB superstring action.

\subsection{The Yang-Mills dielectric matrix model}

Let us consider the matrix model defined by the action
\begin{equation}
S_\epsilon=-\frac{\alpha}{4} \tr \left(\frac{1}{Y}[A_\mu,A_\nu]^2\right)
+V(Y)-\frac{\alpha}{2} \tr \left(\bar{\psi}\Gamma^\mu[A_\mu,\psi]
\right) ,
\label{p1}
\end{equation}
where $Y$ is a new $n\times n$
Hermitian matrix field taken to be {\sl positive definite}, and
$V(Y)$ is a ``potential''. For
reasons which become clear later, the potential is taken to be
\begin{equation}
V(Y)=\beta~{\rm Tr}~Y+\gamma~{\rm Tr}~\ln Y.
\label{p2}
\end{equation}
The partition function is then given by the functional integrals%
\footnote{In order to have a non-trivial saddle point for
$n\rightarrow\infty$, we need to assume that the positive
constants $\alpha$ and $\beta$ are of order $n$.}
\begin{equation}
Z_\epsilon=\int {\cal D}A_\mu ~{\cal D}\psi~{\cal D}Y~e^{-S_\epsilon}.
\label{p3}
\end{equation}
As discussed in the appendix the action is invariant under
\begin{eqnarray}
\delta^{(1)}\psi&=&\frac{i}{4}\{ Y^{-1},[A_\mu,A_\nu]\}~
\Gamma^{\mu\nu}\epsilon,\nonumber \\
\delta^{(1)}A_\mu&=&i\bar{\epsilon}~\Gamma_\mu\psi,\nonumber \\
\delta^{(2)}\psi&=&\xi,\nonumber \\
\delta^{(2)}A_\mu&=&0
\label{p4}
\end{eqnarray}
in the limit $n\rightarrow\infty$. The field $Y$ is assumed to be
invariant with respect to this transformation.

The action~\rf{p1} differs from its classical counterpart~\rf{Yaction}
by the second term on the right-hand side of \eq{p2}. We can e.g. associate
this term with the measure for integration over $Y$ rather
than with the classical action. The classical action~\rf{Yaction}
can be obtained from~\rf{p1} in the limit $\alpha\sim\beta\sim\infty$,
$\alpha/\beta \sim 1$. This limit is associated with
vanishing string coupling constant since~\cite{ikkt}
$\alpha\sim\beta\sim g_s^{-1}$, \ie with the usual classical limit
in string theory.
The matrix model with the action (\ref{p1}) can be considered as the large $n$
reduced model for a ten-dimensional non-abelian ``dielectric'' theory
of the type introduced by 't Hooft \cite{'t Hooft} several years ago for the
abelian case. In this picture the quantity $1/Y$ is the dielectric function
$\epsilon (Y)$, which is then governed by the potential $V(Y)$. Although the
present matrix model and its non-reduced counterpart are much more complicated
than the abelian version\footnote{This applies to the physics as well as to
the mathematics.}, we shall call the model given by eq.~(\ref{p1})
the Yang-Mills dielectric matrix model. This is the reason for the notation
$S_\epsilon$ for the action. Alternatively, one could interpret the field
$Y$ as a rather rudimentary metric.

We start by doing the $Y-$integral. Since the $\psi-$dependent term in
\eq{p1} is independent of $Y$, it is sufficient to consider the integral
\begin{equation}
{\cal F}(z)=\int {\cal D}Y~\exp \left(-\frac{\alpha}{4}~{\rm Tr}
\left(\frac{1}{Y}z^2\right)-\beta~{\rm Tr}Y-\gamma~{\rm Tr}~\ln Y\right).
\label{p5}
\end{equation}
Here $z^2=-[A_\mu,A_\nu]^2$. The integration over the ``angular'' variables
in \eq{p5} is of the Itzykson-Zuber type \cite{it}, and we therefore
get
\begin{equation}
{\cal F}(z)=n!~\alpha^{n(n-1)/2}\prod_{p=1}^{n-1}p!~
\prod_{i=1}^{n}\int_0^\infty dy_i~\frac{\Delta^2(y)}{\Delta(1/y)
\Delta(z^2)}~{\rm exp}\left(-\alpha\sum_i z_i^2/4y_i-\beta\sum_i y_i-\gamma
\sum_i\ln y_i\right),
\label{p6}
\end{equation}
where $\Delta$ is the Vandermonde determinant
\begin{equation}
\Delta(x)=\prod_{i> j}(x_i-x_j)=\det\limits_{ki} x_i^{k-1}.
\label{p7}
\end{equation}
In eq.~(\ref{p6}) the quantities $z_i^2$ and $y_i$ are the eigenvalues of
$-[A_\mu,A_\nu]^2$ and $Y$, respectively. The unitary matrix which
diagonalizes $Y$ has thus been integrated over as in the Itzykson-Zuber
paper \cite{it}.

Now \eq{p6} can be written
\begin{eqnarray}
\Delta (z^2)~{\cal F}(z)&=&n!~\alpha^{n(n-1)/2}\prod_{p=1}^{n-1}p!~
\prod_{i=1}^n\int_0^\infty dy_i~
y_i^{n-1}\prod_{i>j}(y_i-y_j)\nonumber \\
&&\times~\exp \left(-\alpha~\sum_iz_i^2/4y_i
-\beta~\sum_iy_i-\gamma~\sum_i\ln y_i\right).
\label{p8}
\end{eqnarray}
We shall now choose $\gamma$ in such a way that the result of the
$y_i-$integrations gives a result which is ``as string-like as is possible''.
As we shall soon see, this amounts to taking
\begin{equation}
\gamma=n-1/2.
\label{p9}
\end{equation}
Using the well known Bessel integral (from an integral representation of
$K_{-1/2}$ and the explicit formula for this function)
\begin{equation}
\int_0^\infty \frac{dy}{y^{1/2}}~e^{-\alpha z^2/4y-\beta y}
=\sqrt{\pi/\beta}~e^{-\sqrt{\alpha\beta}~z},
\label{p10}
\end{equation}
we obtain by use of (\ref{p9})
\begin{eqnarray}
\Delta(z^2){\cal F}(z)
&=& n!\; \alpha^{n(n-1)/2}\prod_1^{n-1} p!
\;\det\limits_{ki}
\int_0^\infty\frac{dy}{y^{1/2}}\,y^{k-1}
\exp\left(-\frac{\alpha z_i^2}{4y}-\beta y\right)
\nonumber\\*
&=&
n!\; \alpha^{n(n-1)/2}\prod_1^{n-1} p!\;\det\limits_{ki}
(-1)^{k-1}\frac{\partial}
{\partial\beta^{k-1}}
\left[ \sqrt{\pi/\beta}
\exp\left(-\sqrt{\alpha\beta}z_i\right)\right].
\label{p12}
\end{eqnarray}
The matrix whose determinant is to be calculated reads
\begin{equation}
{\cal A}_{ki}\equiv
(-1)^{k-1}~\frac{\partial}
{\partial\beta^{k-1}}
\left( \sqrt{\pi/\beta}
\e^{-\sqrt{\alpha\beta}z_i}\right),
\end{equation}
or, explicitly,
\bea
{\cal A}_{1i}&=&
 \sqrt{\pi/\beta}
\e^{-\sqrt{\alpha\beta}z_i}, \non
{\cal A}_{2i}&= &
\frac{1}{2\beta}\,
 \sqrt{\pi\alpha}z_i
\e^{-\sqrt{\alpha\beta}z_i}
+\frac{1}{2\beta}\,
 \sqrt{\frac{\pi}{\beta}}
\e^{-\sqrt{\alpha\beta}z_i}, \non
{\cal A}_{3i}&= &
\frac{1}{4\beta}\,\alpha\sqrt{\frac{\pi}{\beta}}~z_i^2
\e^{-\sqrt{\alpha\beta}z_i}+\frac{3}{4}\frac{\sqrt{\pi\alpha}}{\beta^2}z_i
\e^{-\sqrt{\alpha\beta}z_i}+\frac{3}{4\beta^2}\,\sqrt{\frac{\pi}{\beta}}
\e^{-\sqrt{\alpha\beta}z_i}, \non
&\ldots&
\eea
The second term in the expression for ${\cal A}_{2i}$ is proportional to the
first line of the matrix, ${\cal A}_{1i}$, and can be omitted in the
determinant. The same property holds for all lines of ${\cal A}$ -- only
the result of the differentiation of the exponential survives in the
determinant, while the terms coming from differentiations of the various
pre-exponential factors are linear combinations of the
previous lines of the matrix. These factors include e.g. the term in
${\cal A}_{3i}$ which is linear in $z_i$. This term is proportional to
the first term in ${\cal A}_{2i}$, etc. etc.

Hence,
\begin{eqnarray}
\Delta(z^2){\cal F}(z)
&=&\tilde{ C}\,\det\limits_{ki}\left[
\sqrt{\pi/\beta} ~\left(\sqrt{\alpha /\beta}~z_i/2\right)^{k-1}
\exp\left(-\sqrt{\alpha\beta}z_i\right)\right] \non
&=& C\,
\Delta(z)
\exp\left(-\sqrt{\alpha\beta}\sum_iz_i\right),
\label{z}
\end{eqnarray}
where the constants $C$ and $\tilde C$ are given by
\begin{equation}
\tilde{C}=n!~\alpha^{n(n-1)/2}\prod_{p=1}^{n-1}p!~~{\rm and}~~
C=n!~((\alpha)^{3/2}/2\sqrt{\beta})^{n(n-1)/2}(\pi/\beta)^{n/2}
\prod_{p=1}^{n-1}p! \:.
\label{p13}
\end{equation}
In these equations, $z_i$ always means the positive square root of the
positive quantity $z_i^2$.

Eq.~(\ref{z}) is one of the main results of this subsection. It shows that
after the {\sl exact} integration over $Y$, for {\sl any} $n$, the action
becomes linear in the
variables $z_i$, although the original action $S_\epsilon$ in eq.~(\ref{p1})
is quadratic in these variables. To the best of our knowledge, this result
is a new matrix model result, and it may therefore be of interest also
outside the present framework. Expressed in terms of string language, we shall
find that we have the option of obtaining the (supersymmetric) Nambu-Goto
type of action, as we shall discuss shortly. The result (\ref{z}) is
possible because of the choice (\ref{p9}) of the power $\gamma$, which allows
us to use the well known explicit (exponential) form for the Bessel
function $K_{-1/2}$. However, it should be noticed that we still
have the ratio of the Vandermonde determinants in eq.~(\ref{p12}),
\begin{equation}
\Delta(z)/\Delta(z^2)=1/\prod_{i>j}(z_i+z_j).
\label{p14}
\end{equation}

To proceed from \eq{p12} we now show that we have the identification
\begin{equation}
\sum_i z_i=\tr\sqrt{-[A_\mu,A_\nu]^2},
\label{p15}
\end{equation}
where it should be remembered that the $z_i^2$'s are the positive eigenvalues
of $-[A_\mu,A_\nu]^2$. Following Dirac we define the square root of a matrix
by its formal power series. Thus we write%
\footnote{
The square root of a matrix can alternatively be defined by the
integral representation
$$
\sqrt{-[A_\mu,A_\nu]^2}=\frac{1}{4\sqrt{\pi }i}\,
\int_{-\infty }^{(0+)}dt\,t^{-3/2}\e^{-t\,[A_\mu,A_\nu]^2},
$$
where the contour of integration encircles counterclockwise the negative
real axis which provides convergence.
Thus we write
$$
Q\equiv\sum_iz_i=\sum_i\sqrt{z_i^2}=
\frac{1}{4\sqrt{\pi }i}\,
\int_{-\infty }^{(0+)}dt\,t^{-3/2}
\sum_{i}\e^{tz_i^2},
$$
which can be represented as
$$
Q=\frac{1}{4\sqrt{\pi }i}\,
\int_{-\infty }^{(0+)}dt\,t^{-3/2}\,{\rm Tr}\e^{-t\,[A_\mu,A_\nu]^2}
={\rm Tr}\sqrt{-[A_\mu,A_\nu]^2}.
$$
All the formulas here and below
can be rigorously derived using this representation.}
\begin{equation}
Q\equiv\sum_iz_i=\sum_i\sqrt{z_i^2}=\sum_i\sqrt{1+(z_i^2-1)}=
\sum_i \sum_{p=0,l\leq p}^\infty \left(\begin{array}{c} p \\ l\end{array}
\right)c_p (-1)^{p-l}z_i^{2l},
\label{p16}
\end{equation}
where $c_p=1\cdot 3\cdot 5\cdot...\cdot (2p-3)/2\cdot 4\cdot 6\cdot \ldots
\cdot 2p$. Now let $U$ be the unitary matrix which diagonalizes the
square of the commutator,
\begin{equation}
{\rm diag}(z_1^2,z_2^2,...,z_n^2)=-U[A_\mu,A_\nu]^2U^\dagger.
\label{p17}
\end{equation}
Then we can write in \eq{p16}
\begin{eqnarray}
Q&=&\sum_{p,l} \left(\begin{array}{c} p \\ l\end{array}\right)(-1)^{p} c_p
\tr\left(U[A_{\mu_1},A_{\nu_1}]^2
U^\dagger U...U[A_{\mu_l},A_{\nu_l}]^2U^\dagger\right)\nonumber \\
&=&\tr\sum_p c_p~\left(-[A_\mu,A_\nu]^2-1\right)^p
=\tr \sqrt{1+(-[A_\mu,A_\nu]^2-1)}
\non
&=&\tr \sqrt{-[A_\mu,A_\nu]^2}.
\label{p18}
\end{eqnarray}
This verifies the identification (\ref{p15}).

By means of the result displayed in eq.~(\ref{p18}), we can rewrite
the partition function (\ref{p3}) by use of eqs.~(\ref{p5}) and (\ref{z})
in the following way
\begin{eqnarray}
Z_\epsilon&=&\int {\cal D}A_\mu{\cal D}\psi{\cal D}Y
\exp \left(\frac{\alpha}{4} \tr \left(\frac{1}{Y}[A_\mu,A_\nu]^2\right)
-\beta\tr Y \right. \non & & ~~~~~~~~~~~~~~~~~~~~
\left.-(n-\frac{1}{2})\tr\ln Y+\frac{\alpha}{2} \tr
\left(\bar{\psi}\Gamma^\mu[A_\mu,\psi]
\right)\right)\non
&=&C\int \left[{\cal D}A_\mu{\cal D}\psi\right]
\exp \left(-\sqrt{\alpha\beta}\tr\sqrt{-[A_\mu,A_\nu]^2}+\frac{\alpha}
{2}\tr(\bar{\psi}\Gamma^\mu[A_\mu,\psi])\right),
\label{p19}
\end{eqnarray}
where the measure in the last expression is defined by
\begin{equation}
\left[{\cal D}A_\mu{\cal D}\psi\right]=\frac{{\cal D}A_\mu{\cal D}\psi}
{\prod_{i>j}(z_i+z_j)}.
\label{measure}
\end{equation}
The result (\ref{p19}) shows that the ``dielectric'' action $S_\epsilon$ is
equivalent to an action which can be considered as a strong coupling
non-abelian Born-Infeld model\footnote{The reader should recall that the
abelian Born-Infeld action \cite{born} has the Lagrangian ${\cal L}=-c\sqrt{1+
F_{\mu\nu}^2/2c}$, where $c$ is a constant. In the strong field limit
${\cal L}$ behaves like $-\sqrt{c/2}\sqrt{F_{\mu\nu}^2}$. It was suggested
many years ago that the strong field limit of the abelian Born-Infeld
should give a field-theoretic description of strings~\cite{nbi}.}.
The new action in (\ref{p19}) is no longer invariant under the
transformations (\ref{p4}). The first of these should be replaced by
\begin{equation}
\delta^{(1)}\psi=\frac{i}{4}\{(-[A_\alpha,A_\beta]^2)^{-1/2},[A_\mu,A_\nu]\}
\Gamma^{\mu\nu}\epsilon.
\label{transf}
\end{equation}
The other transformations in eq.~(\ref{p4}) are unchanged. Thus the effect of
performing the $Y$-integration is not only to produce a new action, but it
also produces a new measure (\ref{measure}) and a new transformation
property (\ref{transf}).

The anti-commutator in eq.~(\ref{transf}) is superfluous, since any operator
commutes with its own square. However, it follows from the appendix that
the action in the second equation (\ref{p19}) is only invariant under the
transformation (\ref{transf}) in the limit $n\rightarrow\infty$. This is
because the last two terms on the right hand side of eq.~(\ref{a3}) do
not vanish when $Y^{-1}$ is replaced by the inverse square root, as in
(\ref{transf}).

\subsection{Connection between the dielectric matrix model and
superstrings in the large {\sl n} limit}

In this section we shall discuss the connection between the super-
``dielectric'' matrix model and superstrings. The possible connection between
the large $n$ limit and strings have been discussed by many authors. We shall
use in particular the QCD-discussion by Bars in ref.~\cite{bars}, where further
references can be found\footnote{We mention some of these works in
\cite{complain}.}. The starting point is an expansion of the
$A_\mu$-field in terms of SU($n$)-generators $l_{\bf k}, {\bf k}=(k_1,k_2)$,
\begin{equation}
(A_\mu)^j_i=C_1\sum_{\bf k}a^{\bf k}_\mu~(l_{\bf k})^j_i,
\label{p20}
\end{equation}
where $a^{\bf k}_\mu$ are expansion coefficients to be integrated over in
functional integrals, and where $C_1$ is some normalization constant.
The expansion (\ref{p20}) is to be compared to the corresponding
expansion of the string variable $X_\mu(\sigma,\tau)$, where $\sigma$
and $\tau$ are the usual world sheet variables. In this case we have
\begin{equation}
X_\mu(\sigma,\tau)=C_2\sum_{\bf k}a^{\bf k}_\mu~e^{i{\bf k\sigma}},
\label{p21}
\end{equation}
where ${\bf \sigma}=(\sigma,\tau)$. The $l$-matrices can be constructed
explicitly in terms of Weyl matrices~\cite{bars}, and they satisfy the
commutation relation
\begin{equation}
[l_{{\bf k}_1},l_{{\bf k}_2}]=i\frac{n}{2\pi}\sin (2\pi{\bf k}_1\times{\bf k}_2
/n)~l_{{\bf k}_1+{\bf k}_2}\rightarrow i({{\bf k}_1}\times{{\bf k}_2})~
l_{{\bf k}_1+{\bf k}_2},~{\rm for}~n\rightarrow \infty,
\label{p22}
\end{equation}
where ${\bf k}_1\times{\bf k}_2={(k_1)}_1{({k_2})}_2-{(k_2)}_1{(k_1)}_2$. In
the following we shall always perform the $n\rightarrow\infty$ limit in the
way done in eq.~(\ref{p22}), even inside sums over the ${\bf k}$'s. We do
not know a rigorous justification of this. Obviously there is a good
chance for the validity of this approach if infinitely high modes ($n
\rightarrow \infty$) are not dynamically relevant. For the bosonic string
this is probably not true, since this string oscillates infinitely much
at short distances due to the tachyon instability. However, the assumption
may be correct for stable superstrings.

The results for SU($n$) mentioned above can be compared on the torus with the
area preserving diffeomorphism. This is discussed in the paper by Bars
\cite{bars} (for a  discussion of SU($n$) on the sphere, see Floratos,
Iliopoulos, and Tiktopoulos \cite{iliaden}), and we shall not repeat this
discussion. We only wish to mention that this approach corresponds to taking
into account the local subalgebra, but ignoring the global translation
generators. The central extensions are thus ignored.

Let us again consider the quantity $Q$ defined in eq.~(\ref{p16}). Inserting
the expansion (\ref{p20}), we get from eq.~(\ref{p18})
\begin{eqnarray}
Q&=&{\rm Tr}\sum_{p,l} \left(\begin{array}{c} p\\l \end{array}\right)(-1)^{p}
c_p\left[-C_1^4\sum_{\bf nmrs}a_\mu^{\bf n}a_\nu^{\bf m}a_\mu^{\bf r}
a_\nu^{\bf s}
({\bf n}\times{\bf m})({\bf r}\times{\bf s})~l_{{\bf n}+{\bf m}}
l_{{\bf r}+{\bf s}}\right]^l\nonumber \\&=&
\sum_{p,l} \left(\begin{array}{c} p\\l \end{array}\right)(-1)^{p-l}c_p~C_1^{4l}
\sum_{{\bf nmrs}}~\prod_{i=1}^l a_{\mu_i}^
{{\bf n}_i}a_{\nu_i}^{{\bf m}_i}a_{\mu_i}^{{\bf r}_i}a_{\nu_i}^{{\bf s}_i}~
({\bf n}\times{\bf m})({\bf r}\times{\bf s})~\nonumber \\
&&\times{\rm Tr}\prod_{j=1}^l
l_{{\bf n}_j+{\bf m}_j}l_{{\bf r}_j+{\bf s}_j}.
\label{p23}
\end{eqnarray}
In the limit $n\rightarrow \infty$ the trace of a product of $l$'s with
different indices produce a Kronecker delta\footnote {This follows from
$$
{\rm Tr}~l_{\bf m}l_{\bf r}=(n^3/(4\pi)^2)~\delta_{{\bf m+r,0}}
$$
and repeated applications of the relation $l_{\bf m}l_{\bf r}=
(n/4\pi)\exp(2\pi i({\bf m\times\bf r})/n)~l_{\bf m+r}$. The exponential
factor produces one plus terms of higher order in $1/n$. Thus,
$$
{\rm Tr}~l_{{\bf m}_1}...l_{{\bf m}_s}=
n(n/4\pi)^s\delta_{{\bf m}_1+...+{\bf m}_s,0}
$$
to leading order.} (up to a normalization factor).
Therefore eq.~(\ref{p23}) becomes
\begin{eqnarray}
Q&=&
n\sum_{p,l}\left(\begin{array}{c} p\\l \end{array} \right)(-1)^{p-l} c_p
\left(\frac{n}{4\pi}C_1^2\right)^{2l}
\sum_{{\bf nmrs}}~\prod_{i=1}^l\left
(a_{\mu_i}^{{\bf n}_i}a_{\nu_i}^{{\bf m}_i}a_{\mu_i}^{{\bf r}_i}a_{\nu_i}^
{{\bf s}_i}~({\bf n}\times{\bf m})({\bf r}\times{\bf s})\right)\nonumber \\
&&\times~\delta_{{\bf n}_1+{\bf m}_1+{\bf r}_1+{\bf s}_1...+{\bf n}_l+
{\bf m}_l+{\bf r}_l+{\bf s}_l,{\bf 0}}.
\label{p24}
\end{eqnarray}
It should be emphasized that this simple result is valid only because
the $n\rightarrow\infty$ limit is taken. Otherwise the trace in eq.~(\ref{p23})
yields a more complicated result than displayed above.
This expression can be rewritten by use of the expansion (\ref{p21}) for
the string coordinates,
\begin{eqnarray}
Q&=&\frac{n}{(2\pi)^2}~\int_0^{2\pi}d^2\sigma \sum_{p,l}
\left(\begin{array}{c} p\\l \end{array}\right)(-1)^{p-l}c_p
\left((n/4\pi)^2(C_1/C_2)^4~\{X^\mu
(\sigma),X^\nu(\sigma)\}_{PB}^2\right)^l\nonumber \\
&=&\frac{n}{(2\pi)^2}~\int_0^{2\pi} d^2\sigma \sqrt{1+
((n/4\pi)^2(C_1/C_2)^4~\{X^\mu(\sigma),X^\nu(\sigma)\}_{PB}^2-1)}\nonumber \\
&=&\frac{1}{2(2\pi)^3}\left(\frac{nC_1}{C_2}\right)^2\int_0^{2\pi} d^2\sigma~
\sqrt{\{X^\mu(\sigma),X^\nu(\sigma)\}_{PB}^2}.
\label{p25}
\end{eqnarray}
Here $\{a,b\}_{PB}$ is the usual Poisson bracket of $a$ and $b$.

Inserting this result in eq.~(\ref{p19}) we finally obtain
\begin{equation}
Z_\epsilon\ra
C\int [{\cal D}X_\mu{\cal D}\psi]\exp\left[-\int d^2\sigma\left(
\sqrt{\alpha\beta}\sqrt{\{X_\mu(\sigma),X_\nu(\sigma)\}_{PB}^2}-\frac{i\alpha}
{2}\epsilon^{ab}\partial_aX^\mu\bar{\psi}\Gamma_\mu\partial_b
\psi\right)\right].
\label{p26}
\end{equation}
The last term in the action follows from expanding $\psi$ in a form similar
to (\ref{p20}) and using that for $n\rightarrow \infty$
\begin{eqnarray}
{\rm Tr}\bar{\psi}\Gamma^\mu[A_\mu,\psi]&=&\sum_{\bf nmr}a_\mu^{\bf n}
\bar{\psi}^{\bf r}\Gamma^\mu\psi^{\bf m}~({\bf n}\times{\bf m})~
{\rm Tr}l_{\bf r}l_{\bf n}l_{\bf m}\nonumber \\
&=&i\frac{n^4C_1C_{\psi}^2}{8(2\pi)^5 C_2}\int d^2\sigma \epsilon^{ab}
\partial_aX^\mu\bar{\psi}\Gamma_\mu\partial_b\psi,
\label{p27}
\end{eqnarray}
where $C_{\psi}$ is the relative normalization of the $\psi$-fields.
The coefficients in (\ref{p25}) and (\ref{p27}) have been absorbed in a
redefinition of $\alpha$ and $\beta$ in the result (\ref{p26}). Thus, e.g.
\begin{equation}
(nC_1/C_2)^2/2(2\pi)^3\sqrt{\alpha\beta}\rightarrow\sqrt{\alpha\beta}.
\end{equation}
It should be emphasized that the measure in the result (\ref{p26}) is
defined through (\ref{measure}). Also, the transformation property
(\ref{transf}) must be replaced by
\begin{equation}
\delta^{(1)}\psi=\frac{1}{2\sqrt{\{X_\alpha,X_\beta\}_{PB}^2}}
\{X_\mu,X_\nu\}_{PB}\Gamma^{\mu\nu}\epsilon
\label{p28}
\end{equation}
in order to ensure invariance of the action in (\ref{p26}).

\newsection{Discussion of the results}

The main result from the dielectric matrix model is given by
eq.~(\ref{p26}). We shall now discuss this result. First, one might wonder if
this formula cannot be applied with $\psi\equiv 0$, so that the bosonic
string would emerge from the dielectric model with $\psi=0$. As already
discussed in connection with eq.~(\ref{p22}) this is highly unlikely, since
we perform the limit $n\rightarrow\infty$ inside sums, like e.g. in the
transition from (\ref{p23}) to (\ref{p24}). This is allowed if the infinitely
high modes are not dynamically relevant. However, we know that due to the
tachyon, at a finite distance of the order the square root of the string
tension, the
bosonic string becomes unstable, due to the relevance of infinitely high modes.
These causes the area of the world sheet to become infinite, due to
an infinitely oscillating string. Hence it is not permitted to interchange
the sum over modes and the limit $n\rightarrow\infty$. For superstrings, the
situation is much more hopeful, since it is stable without a tachyon.
So although we do not have a mathematical proof that
the summation over modes can be interchanged with the limit $n$ goes to
infinity, there are physical reasons to believe that this is possible
for superstrings.

The comparison of the results
 obtained from the matrix model with the ones from superstring
 theory provides a check to what
 extent the superstring theory can be described by the matrix model.
 The calculation of the interaction between D-branes
 is one of such checks.
 It is natural to think that such calculations in the matrix
 model correspond to loop expansion around certain large-$n$
 classical solutions. The one-loop calculations of
 Sect.~2 are performed without summation over $n$, as proposed
 in~\cite{ikkt}, or without integration over $Y$, as proposed
 in the previous section.

 It is worth mentioning, however, that the classical
 solutions~\rf{Dp}, which are associated with D-brane configurations,
 are also classical solutions to the NBI-type matrix model.
 The reason is that these classical solutions are BPS states
 and the commutator $[A_\mu,A_\nu]$ is proportional for them
 to the unit matrix. The same is true for the classical value of $Y$, as
 it follows from eq.~\rf{Ycl}, so the classical
 equations of motion of the NBI model:
\be
\left[ A^{\mu},\left\{Y^{-1},
\left[ A_\mu,A_\nu \right]\right\}\right]=0\,,~~~~~
\left[ A_\mu\,, (\Gamma^\mu\psi)_\alpha \right] =0\,,
\label{ceY}
\ee
are also satisfied.

 A more general property holds in the large--$n$ limit
when any classical solution of the IKKT model is simultaneously
a solution of the classical equations of motion of the NBI model.
To show this, let us rewrite the equations of motion~\rf{ceY} and
 \rf{Ycl} for bosonic matrices in the form
\be
\left\{[ A^{\mu},Y^{-1}],
\left[ A_\mu,A_\nu \right]\right\}
+\left\{Y^{-1},\left[ A^{\mu},
\left[ A_\mu,A_\nu \right]\right]\right\}=0,~~~~
Y^2=-\frac{\alpha}{4\beta}\left[ A_\mu,A_\nu \right]^2.
\label{ceY'}
\ee
 For a solution of the IKKT model, the second term on the left hand side
 of the first equation equals zero. At infinite $n$, when
 the commutators can be replaced by the Poisson brackets,
 the first term also vanishes,
 since the large--$n$ classical equations of motion imply
 $\partial _{\sigma }Y^2=0=\partial _\tau Y^2$~\cite{eguchi} and thus
 the Poisson bracket $\{A_\mu ,Y^{-1}\}_{PB}$ is equal to zero.

In this paper we have not discussed the large $n$ saddle point
configuration of the integral over the matrix field $Y$ in the
partition function $Z_\epsilon$.
It should, however, be emphasized that such a
calculation is very different from the corresponding
``classical'' saddle point calculation, valid for $\alpha\sim\beta
\rightarrow\infty$. In the large-$n$ saddle point, the logarithm of
the Vandermonde determinant enters, and one needs to determine the
spectral density of the eigenvalues $y_i$ which
in turn determines the value of the commutator $[A_\mu,A_\nu]$. This is
most easily seen by summarizing our result in the form
\begin{eqnarray}
Z_\epsilon&=&\int {\cal D}A_\mu{\cal D}\psi{\cal D}Y~\prod_{i>j}(z_i+z_j)
\exp \left(\frac{\alpha}{4} \tr \left(\frac{1}{Y}[A_\mu,A_\nu]^2\right)
-\beta\tr Y \right. \non & & ~~~~~~~~~~~~~~~~~~~~
\left.-(n-\frac{1}{2})\tr\ln Y+\frac{\alpha}{2} \tr
\left(\bar{\psi}\Gamma^\mu[A_\mu,\psi]
\right)\right)\non
&=&C\int {\cal D}X_\mu{\cal D}\psi~\exp\left[-\int d^2\sigma\left(
\sqrt{\alpha\beta}\sqrt{\{X_\mu(\sigma),X_\nu(\sigma)\}_{PB}^2}-\frac{i\alpha}
{2}\epsilon^{ab}\partial_aX^\mu\bar{\psi}\Gamma_\mu\partial_b
\psi\right)\right],
\end{eqnarray}
where now the measure in the last functional integral is the standard one,
whereas the corresponding quantity in the first functional integral is not,
due to the factor $\prod (z_i+z_k)$. This implies that in evaluating the large
$n$ saddle point this additional
$z$ dependent factor in the measure should be taken into account.
Equation~\rf{ceY} for the $A_\mu$-field with a nontrivial
distribution of the eigenvalues of $Y$ possesses undoubtedly a richer
structure than \eq{ce}.

The matrix model given in eqs.~(\ref{p1})--(\ref{p3}) can presumably
be considered as a large $n$ reduced Eguchi-Kawai model for the field
theory with the action
\begin{equation}
S_{\rm field}\propto\int d^{10}x~{\rm Tr}\left(\frac{1}{4}Y^{-1}F_{\mu\nu}^2+
\frac{i}{2}\bar{\psi}\Gamma^\mu D_\mu\psi+V(Y)\right),
\label{p50}
\end{equation}
where $V(Y)$ is the {\sl non-polynomial} potential given in eq.~(\ref{p2}).
As usual, we have
\begin{equation}
F_{\mu\nu}=\partial_\mu A_\nu-\partial_\nu A_\mu+i[A_\mu,A_\nu],~{\rm and}~
D_\mu\psi=\partial_\mu+i[A_\mu,\psi].
\label{p51}
\end{equation}
Because of supersymmetry for $n\rightarrow\infty$, we do not expect that
quenching is necessary~\cite{MK83,ikkt}. Therefore the non-polynomial
action (\ref{p50}), in the limit where $n$ approaches infinity,
could be considered as an effective field theory for superstrings.

{}From this point of view, one could consider the field theory as a regulator
for the Green-Schwarz superstring. For example, this might be useful in
numerical simulations. However, one could also ask if the underlying field
theory could be of direct physical interest. A possible scenario could be
the following: Suppose that at the Planck scale or below there exists
a description in terms of some unified field theory (probably non-polynomial)
with a high order group, like e.g. SU($n$) with $n$ very large. Such a theory
could then, as exemplified by our discussion above, effectively be equivalent
to a Green-Schwarz superstring theory. There would therefore exist a dual
description of the very early universe, either as some unified field theory,
or as a string theory. Of course, it goes without saying that many problems
should be solved, before such a dramatic scenario can be said to be on a
satisfactory basis.

\subsection*{Acknowledgments}

Y.M. and K.Z. are grateful to I. Chepelev for useful discussions.
The work by Y.M. and K.Z. was supported in part by INTAS grant 94--0840,
CRDF grant 96--RP1--253 and RFFI grant 97--02--17927.
Y.M. was sponsored in part by the Danish Natural Science Research Council.
D.S. was funded by the Royal Society.

\setcounter{section}{0}
\setcounter{subsection}{0}
\appendix{On the supersymmetry of the $Y$-integral}

In this appendix we shall show that the action (\ref{p1}) is invariant
under the symmetry transformations (\ref{p4}) in the limit
$n\rightarrow\infty$. If we apply the transformation (\ref{p4}) to the action
(\ref{p1}), we obtain after some calculations
\begin{equation}
\frac{1}{4}~{\rm Tr}\left(\frac{1}{Y}[A_\mu,A_\nu]^2\right)\rightarrow
\frac{i}{2}~{\rm Tr}\left(\epsilon_m\left(\Gamma_0\Gamma_\mu\right)_{mn}\psi_n
[A_\nu,\{[A^\mu,A^\nu],Y^{-1}\}]\right),
\label{a1}
\end{equation}
as well as
\begin{eqnarray}
&&\frac{1}{2}~{\rm Tr}\left(\bar{\psi}\Gamma^\mu[A_\mu,\psi]\right)\rightarrow
\frac{i}{2}~{\rm Tr}\left(\psi_m\left(\Gamma^0\Gamma^\delta\right)_{mp}
\epsilon_p~[A_\nu,\{[A^\nu,A_\delta],Y^{-1}\}]\right)\nonumber \\
&&-\frac{i}{7!~4}\epsilon^{\mu\alpha\beta\lambda_1...\lambda_7}~{\rm Tr}
\left(\psi_m\left(\Gamma^0\Gamma^{11}\Gamma^{\lambda_1}...\Gamma^{\lambda_7}~
\right)_{mp}\epsilon_p~[A_\mu,\{[A_\alpha,A_\beta],Y^{-1}\}]\right).
\label{a2}
\end{eqnarray}
Here the quantity $\{a,b\}$ denotes the anti-commutator of $a$ and $b$, and
should not be confused with the Poisson bracket. In deriving this result we
used the expansion
\begin{equation}
\Gamma^\mu\Gamma^{\alpha\beta}=\eta^{\mu\alpha}\Gamma^\beta-
\eta^{\mu\beta}\Gamma^\alpha-\frac{1}{7!}
\epsilon^{\mu\alpha\beta\lambda_1\ldots\lambda_7}~\Gamma^{11}
\Gamma^{\lambda_1}\ldots\Gamma^{\lambda_7}.
\label{aextra}
\end{equation}
If $Y$ was a c-number, the last term in (\ref{a2}) would vanish for symmetry
reasons, and the first term in this equation always cancel the expression on
the right hand side of (\ref{a1}), corresponding to the well known invariance
of supersymmetric Yang-Mills matrix theory. However, the presence of the
non-commuting $Y$ makes life harder. Here we shall show that the last
term in eq.~(\ref{a2}) vanishes in the limit $n\rightarrow\infty$. Using
the expansion
\begin{eqnarray}
&&[A_\mu,\{[A_\alpha,A_\beta],Y^{-1}\}]=[A_\mu,[A_\alpha,A_\beta]]Y^{-1}
+Y^{-1}[A_\mu,[A_\alpha,A_\beta]]\nonumber \\
&&+[A_\alpha,A_\beta][A_\mu,Y^{-1}]
+[A_\mu,Y^{-1}][A_\alpha,A_\beta],
\label{a3}
\end{eqnarray}
the first two terms give zero contribution when inserted in the last term in
eq.~(\ref{a2}). The critical terms are thus the last two terms on the right
hand side of (\ref{a3}). Consider one of these terms in the large $n$ limit,
\begin{equation}
[A_\alpha,A_\beta][A_\mu,Y^{-1}]=\sum_{\bf mnpr}a_\alpha^{\bf m}a_\beta^{\bf n}
a_\mu^{\bf p}(y^{-1})^{\bf r}\frac{n}{2\pi}\sin\left(\frac{2\pi}{n}{\bf m}
\times{\bf n}\right)\frac{n}{2\pi}\sin\left(\frac{2\pi}{n}{\bf p}\times{\bf r}
\right)~l_{{\bf m}+{\bf n}}l_{{\bf p}+{\bf r}},
\label{a4}
\end{equation}
where we used the expansion (\ref{p20}) and the commutator (\ref{p24}), as well
as an expansion of $Y^{-1}$,
\begin{equation}
(Y^{-1})^i_j=\sum_{\bf r}(y^{-1})^{\bf r}~(l_{\bf r})^i_j.
\label{a5}
\end{equation}
Taking the limit $n\rightarrow\infty$ and using the expansion of $\psi$ we get
\begin{eqnarray}
&&{\rm Tr}(\psi~[A_\alpha,A_\beta][A_\mu,Y^{-1}])\rightarrow\sum_{\bf smnpr}~
\psi^{\bf s}~a_\alpha^{\bf m}~a_\beta^{\bf n}~a_\mu^{\bf p}~(y^{-1})^{\bf r}~
({\bf m}\times{\bf n})~({\bf p}\times{\bf r})~{\rm Tr}(l_{\bf s}
l_{{\bf m}+{\bf n}}l_{{\bf p}+{\bf r}})\nonumber \\
&&\rightarrow \sum_{\bf smnpr}~\psi^{\bf s}~a_\alpha^{\bf m}~a_\beta^{\bf n}~
a_\mu^{\bf p}~(y^{-1})^{\bf r}~({\bf m}\times{\bf n})~({\bf p}\times{\bf r})~
\delta_{\bf s+m+n+p+r,0}\nonumber \\
&&=\int d^2\sigma~\psi~\{X_\alpha (\sigma),X_\beta (\sigma)\}_{PB}~
\{X_\mu (\sigma),1/\sqrt{g(\sigma)}\}_{PB},
\label{a6}
\end{eqnarray}
using the expansion (\ref{p21}). Also, we used
\begin{equation}
1/\sqrt{g(\sigma)}=\sum_{\bf r} (y^{-1})^{\bf r}~e^{i{\bf r\sigma}}.
\label{a7}
\end{equation}
Now we have
\begin{equation}
\epsilon^{\mu\alpha\beta...}~\{X_\mu,1/\sqrt g\}_{PB}\{X_\alpha,X_\beta\}_{PB}
=2\epsilon^{\mu\alpha\beta...} \left(\dot{X}_\mu\dot{X}_\alpha X'_\beta
(1/\sqrt{g})'-X'_\mu X'_\beta\dot{X}_\alpha(1/\sqrt{g})\dot{}\right),
\label{a8}
\end{equation}
where dot and prime denotes derivatives with respect to $\tau$ and $\sigma$,
respectively. The expression on the right hand side of (\ref{a8}) is easily
seen to vanish, since the first term inside the
bracket is symmetric in $\mu$ and $\alpha$, whereas the last term is symmetric
in $\mu$ and $\beta$. Thus, for $n\rightarrow\infty$ we have
\begin{equation}
\epsilon^{\mu\alpha\beta\ldots}~{\rm Tr}(\psi
[A_\alpha,A_\beta][A_\mu,Y^{-1}])
\rightarrow\epsilon^{\mu\alpha\beta\ldots}~\int d^2\sigma~\psi~
\{X_\mu,1/\sqrt g\}_{PB}\{X_\alpha,X_\beta\}_{PB}=0.
\label{a9}
\end{equation}

The proof of the cancellation of the cubic $\psi$ terms,
which emerge in the action (\ref{p1})
under the symmetry transformations (\ref{p4}), is the standard one.

Hence the action (\ref{p1}) is supersymmetric for $n\rightarrow\infty$.

\eop

\end{document}